\pdfoutput=1
\documentclass[english,10pt,aps,prd,a4paper,preprintnumbers,floatfix,nofootinbib,superscriptaddress, twocolumn]{revtex4}
\usepackage[english]{babel}
\usepackage{amsmath}
\usepackage{graphicx}
\usepackage{color}
\usepackage{mathrsfs}   
\usepackage{amssymb}
\usepackage{hyperref}
\usepackage[normalem]{ulem} 
\usepackage{dcolumn}
\usepackage{bm}
\newcommand{\overbar}[1]{\mkern 1.5mu\overline{\mkern-1.5mu#1\mkern-1.5mu}\mkern 1.5mu}
\usepackage{graphicx}
\usepackage[sort&compress]{natbib}

\definecolor{greenLinks}{rgb}{0, 0.6, 0} 
\definecolor{blueLinks}{rgb}{0, 0, 0.6}
\definecolor{redLinks}{rgb}{0.6, 0, 0}
\definecolor{tempText}{rgb}{0.55, 0.10,0.67}
\definecolor{eprintLinks}{rgb}{0.4, 0.4, 0.4}
\definecolor{journalLinks}{rgb}{0.6, 0, 0}

\usepackage{float}

\def\gsim{\raise0.3ex\hbox{$\;>$\kern-0.75em\raise-1.1ex\hbox{$\sim\;$}}}
\def\lsim{\raise0.3ex\hbox{$\;<$\kern-0.75em\raise-1.1ex\hbox{$\sim\;$}}}

\newcommand{\AddrAHEP}{%
  AHEP Group, Institut de F\'{i}sica Corpuscular --
  C.S.I.C./Universitat de Val\`{e}ncia, Parc Cient\'ific de Paterna.\\
 C/ Catedr\'atico Jos\'e Beltr\'an, 2 E-46980 Paterna (Valencia) - SPAIN}

\begin{document}
\preprint{IFIC/18-03}
\title{$\Delta L= 3$ processes: Proton decay and LHC}
\author{Renato M. Fonseca}\email{renato.fonseca@ific.uv.es}
\affiliation{\AddrAHEP}
\author{Martin Hirsch}\email{mahirsch@ific.uv.es}
\affiliation{\AddrAHEP}
\author{Rahul Srivastava}\email{rahulsri@ific.uv.es}
\affiliation{\AddrAHEP}

\begin{abstract}
   \vspace{1cm}

We discuss lepton number violation in three units.  From an effective
field theory point of view, $\Delta L=3$ processes can only arise from
dimension 9 or higher operators. These operators also violate baryon
number, hence many of them will induce proton decay. Given the high
dimensionality of these operators, in order to have a proton half-life
in the observable range, the new physics associated to $\Delta L=3$
processes should be at a scale as low as 1 TeV. This opens up
the possibility of searching for such processes not only in proton
decay experiments but also at the LHC.  In this work we analyze the
relevant $d=9,11,13$ operators which violate lepton number in three
units. We then construct one simple concrete model with interesting
low- and high-energy phenomenology.

\end{abstract}

\maketitle

\section{Introduction}

The standard model conserves baryon ($B$) and lepton ($L$) number
perturbatively. However, this is no longer true for non-renormalizable
operators \cite{Weinberg:1979sa} which might be generated in
ultraviolet completions of the theory. For example, the only dimension
5 ($d=5$) operator, the famous Weinberg operator associated to
Majorana neutrino masses, violates lepton number by two units. On the
other hand, at $d=6$ there are various 4-fermion operators which
violate baryon and lepton number by one unit
\cite{Weinberg:1979sa,Wilczek:1979hc,Abbott:1980zj}, inducing proton
decay to two-body final states.

Proton decay searches therefore concentrate on final states such as
$p\to e^+\pi^0$ or $p\to K^+\bar{\nu}$. The former is expected to be
dominant in ordinary GUTs, while the latter is expected to dominate in
supersymmetric $SU(5)$ models \cite{Nath:2006ut}.  However, all such searches have so
far only provided lower limits on the proton half-life
\cite{Nath:2006ut,Patrignani:2016xqp}.

At the same time, searches for neutrinoless double beta decay have
been negative so far (see for example
\cite{Avignone:2007fu,Deppisch:2012nb} for a review of this
topic). Neither do we have any other clear experimental signal of
lepton number violation.  Thus, we do not know whether neutrinos are
Dirac or Majorana particles.

Given the absence of any experimental signal, it is therefore possible
that some unknown symmetry exists forbidding altogether such
non-renormalizable operators, or perhaps just the lowest order ones.
Such a symmetry could be related to some particular combinations of
lepton/quark flavours, as argued for example in \cite{Hambye:2017qix},
or to total lepton and baryon numbers. The possibility we discuss in
this paper is that lepton number might actually be violated only in
units of three: $\Delta L=3$.

This implies that neutrinos must be Dirac particles, as Majorana mass
terms would require violation of lepton number in two units.  Another
immediate consequence of this hypothesis is that proton decay final
states must be at least three-body, while an unambiguous experimental
signal establishing $\Delta L=3$ requires three charged leptons: $p
\to \pi^-\pi^- e^+e^+e^+$, i.e. a 5-body decay.  It is not hard to see
that $\Delta L=3$ operators must involve at least 3 lepton and 3 quark
fields, which means that they are suppressed by several powers of the
new physics scale.  For such high-dimensional operators, an observable
rate of proton decay is achieved for a new physics scale in the
(1--100) TeV range, depending on the dimension of the operator under
consideration. This also opens up the possibility to actually observe
violation of lepton number in three units at the LHC.  In section
(\ref{sect:model}) we will discuss a concrete model realizing this
idea.

It is worth mentioning that proton decay limits into 4- and 5-body
final states rely on rather old bounds for inclusive decays. Hence,
they would benefit substantially from an up-to-date dedicated
search. In particular there is a lower limit of $0.6 (12) \times
10^{30}$ years on the nucleon lifetime associated with $p/n \to e^+
(\mu^+) $+anything \cite{Patrignani:2016xqp,Learned:1979gp} which
relies on experiments \cite{Reines:1974pb,Cherry:1981uq} done in the
70's and early 80's with exposures three to four orders of magnitude
lower than the one achieved in Super-Kamiokande.  Super-K has recently
published some limits on 3-body decays \cite{Takhistov:2014pfw}:
$\tau\left(p\rightarrow e^{+}\nu\nu\right)>1.7\times10^{32}$ and
$\tau\left(p\rightarrow\mu^{+}\nu\nu\right)>2.2\times10^{32}$ at 90\%
confidence level, but gives no limits on 4-body and 5-body decays. As
neutrinos are missing energy, they may as well be anti-neutrinos,
hence the quoted values apply both to $\Delta L=1$
($p\rightarrow\ell^{+}\overbar{\nu}\nu$) and $\Delta L=3$
($p\rightarrow\ell^{+}\overbar{\nu}\overbar{\nu}$) decay modes.
Nevertheless, we mention that in models with lepton number violation
in three units, the decay channels with more particles in the final
state (4 or more) might be the dominant ones, as we will discuss latter.
 In this case one has to rely on the rather weak inclusive
limits mentioned above.

We would like to mention that we are not the first to discuss $\Delta
L=3$ processes and proton decay into multi-particle final states
(although the literature on this subject seems to be quite scarce). In
\cite{Weinberg:1980bf} an operator analysis of
$\Delta\left(B,L\right)=\left(1,\pm3\right)$ was carried out, while the
authors of \cite{Durieux:2012gj} considered $\Delta
L=3$ LHC processes involving all generations. Three
lepton proton decay modes resulting from $d=9,10$ operators have been
discussed recently in \cite{Hambye:2017qix}.  However, all final
states considered in this paper are $\Delta B=\Delta L=\pm1$. In
\cite{Kovalenko:2002eh} 4-body and 5-body proton decays (all with
$\Delta L=\pm1$) have been discussed in the context of leptoquark (LQ)
models. The only other paper, that we are aware of, that mentions a
$\Delta L=3$ proton decay is \cite{Appelquist:2001mj}. This paper
discusses a $d=17$ operator that appears in an extra-dimensional model
with six spacetime dimensions. Finally, we note that one can also consider the
reverse situation where baryon number is violated in three units, while lepton number
is changed by one unit only \cite{Babu:2003qh}.

The rest of this paper is organized as follows. In section
(\ref{sect:ops}) we discuss $\Delta L=3$ operators. After that, in section
(\ref{sect:model}), we provide one example of a concrete model
which generates $\Delta L=3$ processes observable not only in proton decay experiments,
but possibly also at the LHC. In section (\ref{sect:sym}) we
analyze the possible connection of $\Delta L=3$ processes
with Majorana/Dirac neutrinos. We then close with a short
summary.

\section{ $\Delta L=3$ operators}
\label{sect:ops}

We start with a discussion on the effective operators which lead to
lepton number violation in three units.  It is well known that (a)
baryon number $B$ and lepton number $L$ breaking has to occur in
integer units and (b) $\left(\Delta B,\Delta
L\right)=\left(\textrm{odd},\textrm{odd}\right)$ or
$\left(\textrm{even},\textrm{even}\right)$. The reason why $\Delta B$
must be an integer is simply due to the fact that in order for an
operator to be colorless, the number of quarks (triplets of $SU(3)_C$)
minus the number of anti-quarks (anti-triplets) must be a multiple of
3. On the other hand, if the $\Delta B$ associated to a given operator
is even (odd), then the number of quarks plus anti-quarks is also even
(odd), in which case Lorentz invariance requires the presence of an
even (odd) number of leptons plus anti-leptons, hence $\Delta L$ will
also be even (odd).  Thus we have:
\begin{equation}
\Delta B+\Delta L=0,\pm2,\pm4,\cdots
\end{equation}
As a consequence, lepton number violation in three units is only
possible if baryon number is violated as well. The simplest solution
to consider is $\Delta B=\pm 1$; $\Delta B=\pm 3$ and more complicated
possibilities involve at least 12 fermions (this is the case for some
non-perturbative effects \cite{tHooft:1976rip,
  tHooft:1976snw}). Furthermore, having $\Delta L$ and $\Delta B$ with
the same sign leads to lower-dimensional operators, so in this sense,
beyond the normal $\Delta\left(B,L\right)=(1,\pm1)$ proton decay
operators, the $\Delta\left(B,L\right)=\left(1,3\right)$ operators are
the next simplest ones.

In the absence of a symmetry, one usually expects that normal proton
decay operators dominate over the higher-dimensional ones with
$\Delta\left(B,L\right)=\left(1,\pm3\right)$. Perhaps the simplest way
to explain the non-observation of simpler decay final states is to
invoke the presence of a $Z_3(L)$ symmetry. However, note that one
could as easily eliminate the $\left|\Delta L\right|=1$ operators and
keep the $\Delta L=\pm3$ ones with other discrete $Z_n$ symmetries
involving baryon number as well. In fact, continues symmetry groups
$U(1)_{3B\mp L}$ will work equally well.

Let us consider from now on the sign of $\Delta B$ to be positive. A
$\Delta L=\pm3$ operator necessarily involves at least 6 standard
model fields --- 3 quarks and 3 (anti)leptons --- thus the lowest
order operator must be 9-dimensional \cite{Weinberg:1980bf}.  While
there are dozens of $d=9$ operators, only two of them violate $L$ by
three units ($\Delta L=+3$). These are:
\begin{eqnarray}\label{eq:d9_1}
{\cal O}_9^1 = \overbar{ u^c}\overbar{u^c}\overbar{u^c}\overbar{e^c}LL, \\ \label{eq:d9_2}
{\cal O}_9^2 = \overbar{u^c}\overbar{ u^c}QLLL. 
\end{eqnarray}
However, neither leads to proton decay, since both of these operators
have the $\overbar{u^c}$'s contracted in an anti-symmetric fashion, hence
two up-type quarks of different generations are necessary.\footnote{On
  the other hand, accelerator experiments such as the LHC could probe
  these operators. Note also that the operator in equation
  \eqref{eq:d9_1} requires two different $L$ generations.}

At $d=10$ one finds for the first time a single $\Delta L =-3$
operator \cite{Weinberg:1980bf} that can induce proton (and neutron)
decay,
\begin{equation}
\mathcal{O}_{10}=\overbar{d^{c}}\overbar{d^{c}}\overbar{d^{c}}\overbar{L}\overbar{L}\overbar{L}H^{*}\\
\end{equation}
which is non-zero only for two or more generations of either $d^c$ or
$L$. From electric charge conservation, one can infer that it will
induce processes with two neutrinos, such as $n\to e^- \nu \nu \pi^+$
and $p\to e^- \nu \nu \pi^+ \pi^+$.\footnote{Final states with muons
  instead of electrons, and/or kaons instead of pions, are allowed in
  all cases we discuss in this paper.}

At $d=11$, including derivatives, there are already 14 $\Delta L=3$
operators.\footnote{The counting we present here refers to the number
  of gauge and Lorentz invariant field combinations; it does not take
  into account different contractions of the same fields nor different
  applications of the derivative operator.} All of them contain either
two derivatives, two Higgs fields, or one derivative and one
Higgs. For example:
\begin{eqnarray}\label{eq:d11}\
  {\cal O}_{11}^1 =  \partial\partial\overbar{u^c}\overbar{u^c}QLLL ,\\
  {\cal O}_{11}^2 =
  \partial Q\overbar{u^{c}}\overbar{u^{c}}LL\overbar{e^{c}}H .
\end{eqnarray}
${\cal O}_{11}^1$ induces both proton and neutron decay of the type
$p\to e^+ {\overbar \nu}{\overbar \nu}$, $p \to \pi^- e^+ e^+{\overbar \nu}$ and
$n\to \pi^-\pi^- e^+e^+{\overbar \nu}$.  From the Super-Kamiokande limit
on $p\to \ell^+ \nu \nu$ \cite{Takhistov:2014pfw} we estimate very
roughly a lower limit on the scale of ${\cal O}_{11}^1$ of order
(7--13) TeV for couplings of order ${\cal
  O}(1)$.\footnote{Any realistic ultra-violet completion of this
  operator will, however, have to obey also low-energy bounds on the
  couplings, which might lead to quite stringent limits on some
  couplings, and correspondingly weaker limits on the scale of the
  operator, depending on the model. For a discussion in one concrete
  model see the next section.}
We defer the discussion of ${\cal O}_{11}^2$ to section 
(\ref{sect:model}). 

There are more $\Delta L=-3$ operators at dimension $d=12$ which we
will not discuss in detail. Importantly, all of the operators
discussed so far, with $d\leq12$, involve neutrinos, making it
impossible to tag lepton number experimentally. In other words, it is
not possible to be certain that lepton number is violated in three
units unless all the leptons are charged. One can easily see from
electric charge conservation alone that 8 fermions are needed to form
such an operator:
\begin{equation}
eeeuuuu\overbar{d}\left(H\textrm{ or }\partial\right)\,.
\end{equation}
The need for a derivative or a Higgs boson is seen once the full
standard model group is taken into consideration. There are many
operators at $d=13$ of this type; we show here only
two examples:
\begin{eqnarray}\label{eq:d13_1}
{\cal O}_{13}^1 = \partial \overbar{u^c}\overbar{u^c}\overbar{u^c}\overbar{u^c}d^c
\overbar{e^c}\overbar{e^c}\overbar{e^c} ,
\\ \label{eq:d13_2}
{\cal O}_{13}^2 = \partial \overbar{u^c}\overbar{u^c}d^cQQ\overbar{e^c}LL \,.
\end{eqnarray}
Note that ${\cal O}_{13}^1$ requires two lepton generations, leading
to decays of the form $p\rightarrow e^{+}
e^{+}\mu^{+}\pi^{-}\pi^{-}$. The operator ${\cal O}_{13}^2$, on the
other hand, can yield decays involving just one generation of quarks
and fermions: $p \to e^+e^+e^+ \pi^- \pi^-$, $p \to e^+e^+ {\overbar\nu}
\pi^-$, $p \to e^+ {\overbar\nu} {\overbar\nu} \pi^0$, and also $n \to e^+ e^+
e^+ \pi^- \pi^- \pi^-$, $n \to e^+ e^+ {\overbar\nu} \pi^- \pi^-$, $n \to
e^+ {\overbar\nu} {\overbar\nu} \pi^-$. From simple phase space arguments one
expects the 4-body decay modes to dominate over the other
ones. However, at least in principle, the final state with three
positrons can have a half-life short enough to be observed in (future)
proton decay searches. We will discuss this in the context of one
concrete ultra-violet completion for operator ${\cal O}_{13}^2$ in the
next section.

\section{A simple model for $\Delta L=3$ with charged leptons}
\label{sect:model}

Many ultra violet complete models generating $\Delta L=3$, $d =
13$ operators can be constructed. Here, for illustration and to
facilitate definite quantitative discussions, we will discuss one
simple example. In our model, proton decay is induced by a $d=13$
operator which is generated at tree-level, but the model also provides
a $d=11$ 1-loop contribution to this process.  The $d=11$ loop
contribution, as we will discuss shortly, is suppressed compared to
the $d=13$ tree level contribution which is the dominant contribution
to proton decay. This illustrative model is also chosen for its
minimality of particle content while still being able to potentially
provide a clear $\Delta L = 3$ signal both in proton decay and at the LHC.

In our construction, we make a simple extension of the Standard Model
by adding left-handed fermions $N,N^c \equiv F_{1,1,0}$ and two types
of scalars $S_u \equiv S_{\bar3,1,-2/3}$ and $S_d \equiv
S_{\bar3,1,1/3}$. For reasons discussed below, we will need two copies
of this last field, $S_d$ and $S^\prime_d$, three generations of $N$,
and six of $N^c$.\footnote{A truly minimal setup would postulate only
  three copies of $N^c$ and one of $N$. In this case, two of the $N^c$
  then pair off with the active neutrino of the SM, to generate the
  two mass splittings observed in oscillation experiments, while the
  third $N^c$ forms a vector-like pair with $N$. } Here, $S$ and $F$
stand for scalars and (left-handed) Weyl fermions, and the subscripts
indicate the transformation properties/charges of the fields under the
Standard Model gauge group, following the order $SU(3)_C\times
SU(2)_L\times U(1)_Y$.

The Lagrangian contains the following pieces:
\begin{eqnarray}\label{eq:lag}
  {\cal L } & = & {\cal L }_{SM}+ Y_\nu L N^c H + Y_1 \overbar{u^c} \overbar{N^c} S_u
  + Y_2 N^c d^c S_d^{*}  \nonumber \\ 
  & + &  Y_3 \overbar{e^c} \overbar{u^c} S'_d + Y_4 Q L S_d +  \mu S_u S_d S'_d
  + m_{N} N N^c \nonumber \\
& + & \cdots ,
\end{eqnarray}
where the dots stand for additional terms which are irrelevant to the
following discussion. We do not consider terms such as $N N$,
$N^c N^c$, $L N H$, $\overbar{u^c} \overbar{N} S_u$, $ N d^c
S_d^{*}$, $d^c d^c S_u$, $u^c d^c S_d$, $Q Q S^{*}_d$ and similar terms with $S_d \to S'_d$,
as they are forbidden by $Z_3(L)$ symmetry under which each field has
a charge $\omega^L$, $\omega$ being the cubic root of 1, and $L$ the
field's lepton number. It is straightforward to check that $L\left(S_d
\right)=L\left(S^\prime_d \right)=L\left(S_u \right)=L\left(N^c \right)=-L\left(N \right)=-1$.

The term proportional to $\mu$ couples three coloured triplets, thus
two copies of $S_d$ are needed. In the limit of $\mu\to 0$ the model
conserves both $B$ and $L$ and the proton is completely stable, hence
$\mu$ can be seen as the source of lepton and baryon number
violation. Note that the mass term for the $N$, $N^c$ fermions is of
the Dirac-type, hence it does not violate $L$.

At this point, it is necessary to briefly discuss the need for the
additional $Z_3(L)$ symmetry. As mentioned earlier, the gauge quantum
numbers of the scalar $S_d$ allow a term $Y_5 QQ S^*_d $. This
coupling, however, together with eq. (\ref{eq:lag}), induces proton
decay via a tree-level $d=6$ operator with $\Delta(B+L)= 2,
\Delta(B-L)=0$ at an unacceptable rate. It follows that the product of
$Y_5$ with other couplings must be very small; indeed, we estimate
that\footnote{We assume all couplings to be real without loss of
  generality.}
\begin{equation}\label{eq:cp6}
Y_4 Y_5 \lsim 10^{-24} \Big(\frac{m_{S_d}}{1 \hskip1mm {\rm TeV}}\Big)^2 ,
\end{equation}
and similarly for the product $Y_3 Y_5$. This motivates strongly the
introduction of a symmetry under which $L$ and $S_d/S'_d$ are odd
while Q is even, to forbid the unwanted term. There are different ways
to do this and a simple $Z_3(L)$ or $Z_4(B+L)$ is sufficient (in our model,
both cases lead to an accidental $U(1)_{3B-L}$).

A similar, but quantitatively much less important concern is that
$S_d$/$S_d'$ can have simultaneous coupling to $\overbar{ e^c} \overbar{u^c}$
and $Q L$, since these scalars transform in the same way under all
symmetries. Constraints from meson decays limit the product of these
couplings roughly to \cite{Davidson:1993qk,Dorsner:2016wpm}
\begin{equation}\label{eq:pion}
Y_3 Y_4 \lsim 2 \cdot 10^{-5} \Big(\frac{m_{S_d}}{1 \hskip1mm {\rm TeV}}\Big)^2 .
\end{equation}
Constraints on $Y_3$ and $Y_4$ individually are much weaker. From the
constraints discussed in \cite{Davidson:1993qk,Dorsner:2016wpm}, we
estimate:
\begin{eqnarray}\label{eq:pionGF}
Y_3 \lsim 0.26 \Big(\frac{m_{S_d}}{1 \hskip1mm {\rm TeV}}\Big), \\ \nonumber
Y_4 \lsim 0.27 \Big(\frac{m_{S_d}}{1 \hskip1mm {\rm TeV}}\Big).
\end{eqnarray}
Thus, in order to maximize the proton decay rate it is preferable that
each scalar couples to either $\overbar{e^c} \overbar{u^c}$ or $Q L$, but
not both. This can be achieved by introducing another discrete
symmetry which would eliminate the unwanted couplings, as we will
explain in the next section. Also for simplicity we assume that all of
the lepto-quarks, namely $S_u, S_d$ and $S'_d$, couple preferentially to
first generation quarks and leptons only. However, this requirement is
not essential and can be relaxed.

Additionally, there are constraints from direct searches at the
LHC. For both $S_d$ and $S_d'$, standard lepto-quark searches
apply. Limits on these states depend on the lepton and quark
generations they couple to. We are mostly interested in first
generation lepto-quarks, and for this case, searches from CMS
\cite{CMS:2016imw} and ATLAS \cite{Aaboud:2016qeg} establish a lower
limit of roughly $m_{S_d} \sim m_{S'_d} \sim 1$ TeV. For $S_u$, as
discussed below, in order to have a clear signature of $\Delta L = 3$
process at LHC, final states should always be 4-body, hence
constraints are slightly less stringent, but in any case we expect
them not to be significantly below 1 TeV.

\begin{center}
\begin{figure}[tbph]
\begin{centering}
\includegraphics[scale=0.75]{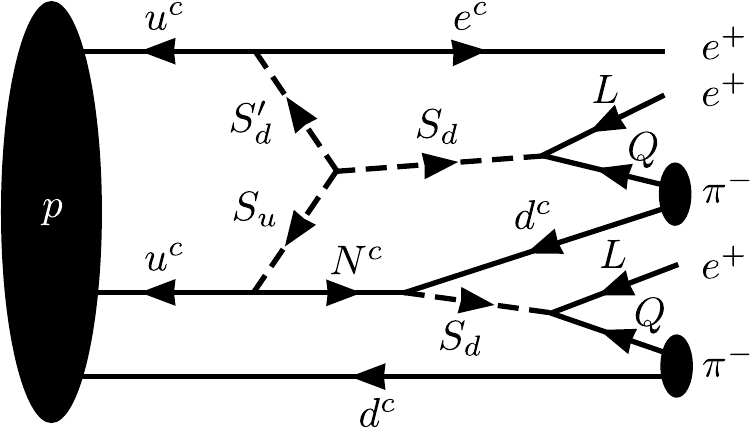}
\end{centering}
\protect\caption{\label{fig:dim13} Proton decay
  $p\rightarrow\pi^{-}\pi^{-}e^{+}e^{+}e^{+}$ induced by a $d=13$
  operator in our model. Diagrams permuting the 4 external up-quarks
  are equally allowed.  Note the presence of the trilinear coupling
  $\mu S_u S_d S^\prime_d$ in the diagram.}
\end{figure}
\par
\end{center}

Fig. (\ref{fig:dim13}) shows the tree-level Feynman diagram for proton
decay in our model. The parton level amplitude is given approximately
by the expression
\begin{eqnarray}\label{eq:amp13}
{\cal A} \sim   
\frac{Y_1 Y_2 Y_3 Y_4^2\mu \left\langle p\right\rangle }
     {m_{N}^2 m_{S_u}^2 m_{S_d}^4 m_{S'_d}^2}\,,
\end{eqnarray}
where $\left\langle p\right\rangle$ is the mean parton momentum involved in
the process. Very roughly $\langle p\rangle \sim {\cal O}(m_p)$.

A simple estimate of the mean proton lifetime associated to this decay mode is given by
\begin{align}\label{eq:pdecay_pipieee_v1}
	\tau^{-1}\left(p\rightarrow 3e^{+}2\pi^{-}\right) &
	\sim \frac{J_0}{f\left(5\right)} \, \mathcal{A}^{2}\frac{m_{p}^{15}W^2}{f_{\pi}^{2}}\,,
\end{align} 
where $f\left(n\right)\equiv4\left(4\pi\right)^{2n-3}\left(n-1\right)!
\left(n-2\right)!$ takes care of the phase space volume available to
the decay products,\footnote{For $n=2,3,4,5$, this factor is close to
  $5\times 10^1$, $2\times 10^4$, $2 \times 10^7$ and $3 \times
  10^{10}$.} $f_\pi=0.13$ GeV is the pion decay constant, $W= \langle \pi^- | \mathcal{O} | p
\rangle\sim0.2\textrm{ GeV}^{2}$ \cite{Aoki:2017puj} is the QCD form factor involved in single pion decay, and
$J_{0}\approx0.1$ is a numerical factor \cite{Claudson:1981gh}. Without taking into account form factors,
one would replace $J_0 W^2/f_\pi^2$ by $m_p^2$ in eq. \eqref{eq:pdecay_pipieee_v1}. 

These two estimates for the lifetime associated to  $p\rightarrow e^{+}
e^{+}\mu^{+}\pi^{-}\pi^{-}$ are shown in fig. (\ref{fig:pdecay}) as two lines which form the upper and lower limits of the blue band. Note that 
due to LHC constraints, none of the scalars can be light. However,
there are practically no constraints on the mass of $N^c$, hence we show
the estimated lifetime as a function of the mass of this field.

By switching the $SU(2)_L$ components of the $Q$ and $L$ fields, it is
possible to use the operator shown in fig. (\ref{fig:dim13}) to induce
the 4-body decays $p\rightarrow\pi^{-}e^{+}e^{+}\overbar{\nu}$ and
$p\rightarrow\pi^{0}e^{+}\overbar{\nu}\overbar{\nu}$ --- see
fig. (\ref{fig:dim13_othermodes}). The associated lifetimes are given
by the expression in eq. \eqref{eq:pdecay_pipieee_v1} but without a
factor $J_0 m_p^2/f_\pi^2$ and with the
larger phase-space factor $1/f(4)$ replacing $1/f(5)$, implying that
the 4-body decay is roughly a factor of $10^3$ faster than the 5-body
one. This is shown in fig.  (\ref{fig:pdecay}) as the upper limit of the orange band,
while the lower limit corresponds to a naive power counting estimate of the the lifetime, obtained by substituting $W^2$ by $m_p^4$.
 The difference between the
proton lifetime in decay modes
$p\rightarrow\pi^{-}e^{+}e^{+}\overbar{\nu}$ and
$p\rightarrow\pi^{0}e^{+}\overbar{\nu}\overbar{\nu}$ is negligible
and so only the latter mode is shown in fig. (\ref{fig:pdecay}).

\begin{center}
\begin{figure}[tbph]
\begin{centering}
\includegraphics[scale=0.75]{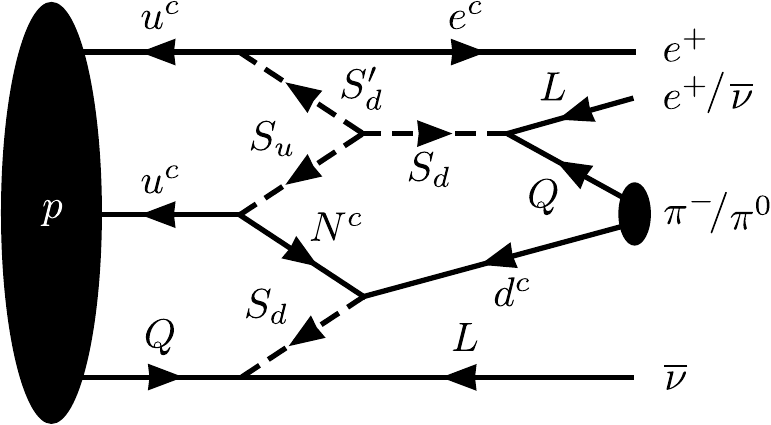}
\end{centering}
\protect\caption{\label{fig:dim13_othermodes} Proton decay into four
  bodies, $p\rightarrow\pi^{-}e^{+}e^{+}\overbar{\nu}$ and
  $p\rightarrow\pi^{0}e^{+}\overbar{\nu}\overbar{\nu}$, induced by
  the same $d=13$ operator as in fig. (\ref{fig:dim13}).}
\end{figure}
\par
\end{center}

As mentioned earlier, there are currently no limits on 4- and 5-body
decays from Super-Kamiokande, hence the current bounds on the
processes $p\rightarrow\pi^{-}\pi^{-}e^{+}e^{+}e^{+}$ and
$p\rightarrow\pi^{0}e^{+}\overbar{\nu}\overbar{\nu}$ are rather weak
($\sim 10^{30}-10^{31}$ years). Hyper-Kamiokande \cite{Abe:2011ts} is
expected to be able to probe 2-body proton decay modes up to $10^{35}
$ years. Thus, one should not exclude the possibility of a five orders
of magnitude improvement in the experimental reach on $\Delta L=3$
decay modes. In this context we mention also the DUNE experiment
\cite{Acciarri:2015uup}, although its mass is smaller than the one of
Hyper-Kamiokande. Note however that neither of these experiments has
charge discrimination, hence they will not be able to unambiguously check that
lepton number is being violated in 3 units.

In our model, the decays involving neutrinos dominate, hence they
should probably be seen first in events with 3 Cherenkov rings (with
at least two being of shower-type) associated to an invariant mass
below $m_p$. However, it is conceivable that Hyper-Kamiokande or DUNE
can also observe the rarer mode with three charged leptons in the
final state, given that this mode involves a clean signal with 3
shower- plus 2 non-shower-type Cherenkov rings, and a reconstructed
mass equal to $m_p$.

\begin{center}
\begin{figure}[tbph]
\begin{centering}
\includegraphics[scale=0.45]{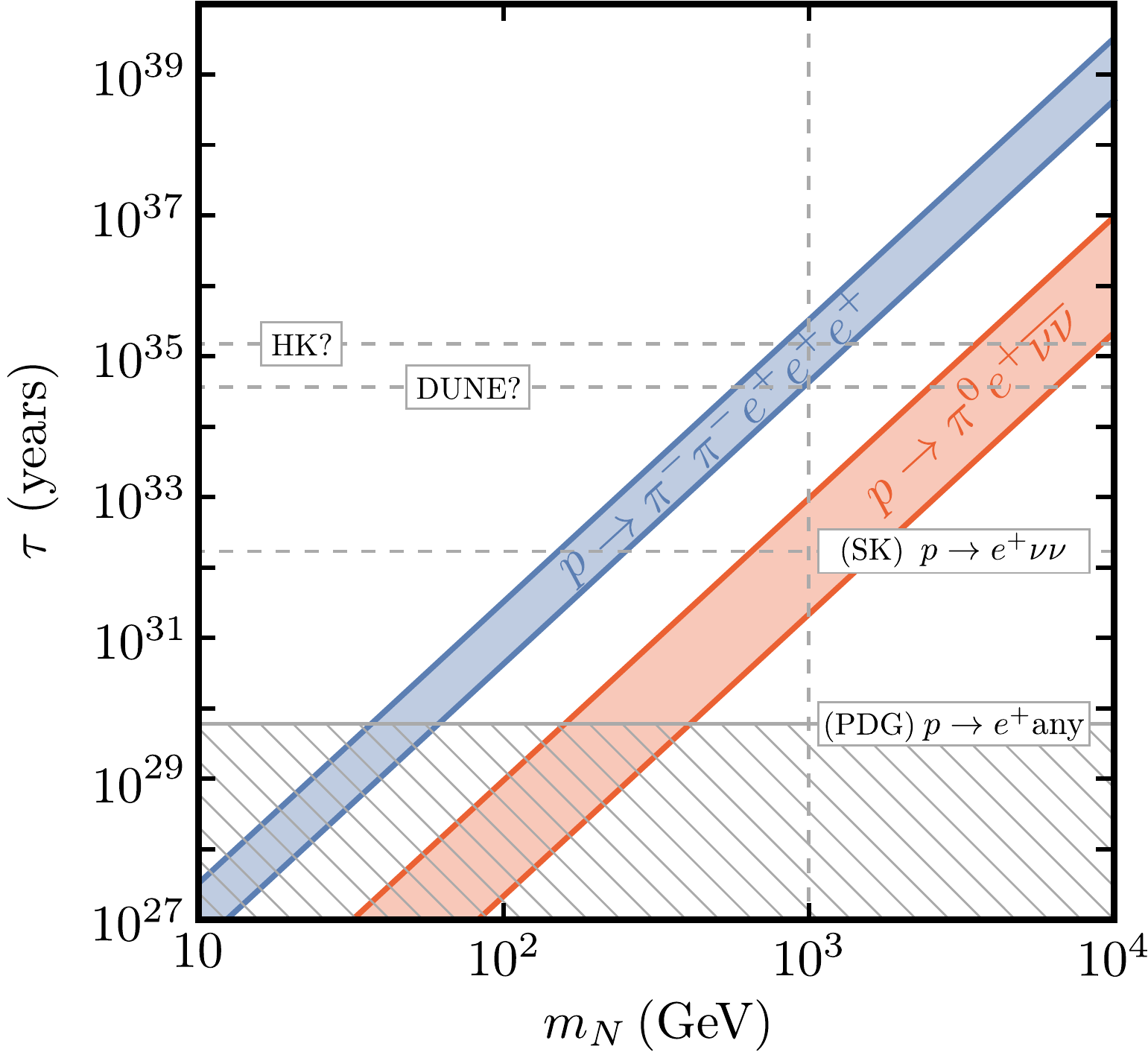}
\end{centering}
\protect\caption{\label{fig:pdecay} Proton decay lifetime as function
  of $m_{N}$, for $\mu=10$ TeV, $m_{S_u}=m_{S_d}=m_{S^\prime_d}=1$
  TeV, $Y_1=Y_2=1$, $Y_3=0.26$ and $Y_4=0.27$. The figure also shows
  existing constraints from proton decay searches, for the chosen
  benchmark point. Inclusive searches exclude $ m_{N}$ smaller than
  $\sim 150$ GeV. Also, for reference, are shown the current
  Super-Kamiokande limits on three body decay mode $p\rightarrow
  e^{+}\overbar{\nu}\overbar{\nu}$, which is currently the most
  stringent limit for a three body decay mode. Note, however, that
  this limit does not apply to our case. We have also shown the
  expected reach of the Hyper-Kamiokande and DUNE experiments for two
  body decay modes.}
\end{figure}
\par
\end{center}

We now turn to a brief discussion of $\Delta L=3$ phenomenology at the
LHC. The new scalars of our model can be produced either in pairs
(through gluon-gluon fusion), in association with $N^c$, or in
association with a Standard Model lepton (in the case of
$S^{(\prime)}_d$).  Production cross sections for these lepto-quark
states at LHC have been calculated several times in the literature,
most recently at next-to-leading order in \cite{Dorsner:2018ynv}, for
example. If the lepto-quark masses are around 1 TeV one expects that
the current run of the LHC will find at least some hints. However,
even for a lepto-quark with 2 TeV of mass, one expects roughly a pair
production cross section of $10^{-2}$ fb, or 30 events in 3000
$\textrm{fb}^{-1}$ before cuts. Thus, LHC signatures for $\Delta L =
3$ processes are possible. Associated production cross sections are
larger for small $N^c$ masses and large couplings, given current
constraints up to $(1-2)$ fb for the fields $S_u$ and $S_d, S'_d$,
assuming couplings equal to $1$. However, associated production cross
sections scale like $Y_i^{ 2}$ and thus could conceivably be much
smaller.

Most interesting for us is $S_u$ production, as it can be used to
experimentally probe $\Delta L=3$ violation at the LHC. The scalar
$S_u$, once produced, has two decay modes which are shown in
fig. (\ref{fig:sudecay}). Given that we know neither the masses nor
the couplings involved, we can not predict which of these two decays
will be dominant. However, in order to observe violation of lepton
number in three units, both channels should have similar branching
ratios for the following reason.  Gluon-gluon fusion
actually produces a pair of scalars $S_u$ and $S_u^{*}$.  Thus, if
one of the two channels in fig. (\ref{fig:sudecay}) dominates, one
either has the final state $2e^+2e^-$ plus jets (through $S_d$) or
$e^+e^-$ plus jets (through the diagram with $N^c$).  If, on the other
hand, one of the two scalars decays through a diagram involving
$S_d$'s while the other decays through the diagram with $N^c$, the
final state can be $3e^+$ plus jets (or $3e^-$ plus jets). One can
easily convince oneself that this gives a sizable event number only if
Br$(S_u \to (S_d^*)^\star + (S^{\prime *}_d)^\star \to 2 e^- + 2 j)$
$\simeq$ Br$(S_u \to (N^{c})^\star + {u^c} \to e^+ + 3 j)$. Here, the
stars `$\star$' indicate that the intermediate state might be off-shell.  Given
the lower limits on the mass of $S_d, S'_d$, this most likely requires
that $N^c$ to be heavy too: $m_{N^c} \gsim m_{S_u}$. If this is not
the case, the 2-body decay Br$(S_u \to N^c + {u^c})$ will dominate
over the channel with $S_d$. However, in this region of parameter
space where $N^c$ is rather heavy, we expect that the proton decay
lifetime into $3e^+2\pi^-$ is rather long --- see
fig. (\ref{fig:pdecay}). Thus, the LHC and proton decay experiments
test complementary parts of the parameter space of the model (heavy
versus light $N^c$).

\begin{center}
\begin{figure}[tbph]
\begin{centering}
\includegraphics[scale=0.75]{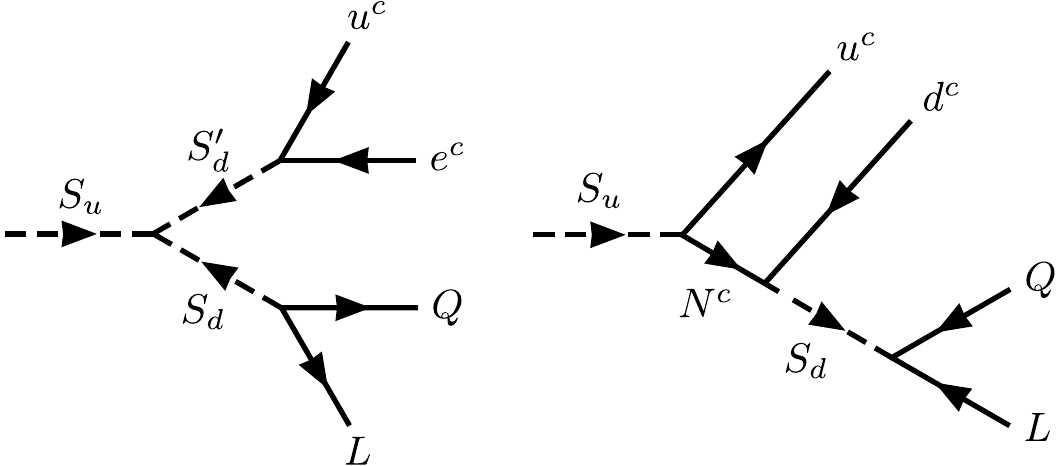}
\end{centering}
\protect\caption{\label{fig:sudecay} Feynman diagrams for $S_u$ decay
  at the LHC.}
	\end{figure}
	\par
\end{center}

The operator discussed so far is 13-dimensional: $\partial
QQ\overbar{u^{c}}\overbar{u^{c}}d^{c}LL\overbar{e^{c}}$. As
mentioned in the introduction, there are lower dimensional operators
which also break lepton number in three units. Hence, they are allowed
by all symmetries of our model. Indeed in our model one can generate
both dimension 9 and 11 operators with $\Delta L=3$; we provide one
example in fig. \eqref{fig:loop_operator}. However, these are
suppressed by the smallness of neutrino masses, the smallness of
first/second generation Yukawa couplings,\footnote{Proton decay only
  involves first generation up quarks and first or second generation
  down quarks and leptons. Note also that the only two $d=9$ operators
  --- see eqs. \eqref{eq:d9_1} and \eqref{eq:d9_2} --- necessarily
  require two up-quark generations.} and/or loop factors. Hence, they
are not as important as the $d=13$ operator in
figs. \eqref{fig:dim13} and \eqref{fig:dim13_othermodes}.

\begin{center}
\begin{figure}[tbph]
\begin{centering}
\includegraphics[scale=0.75]{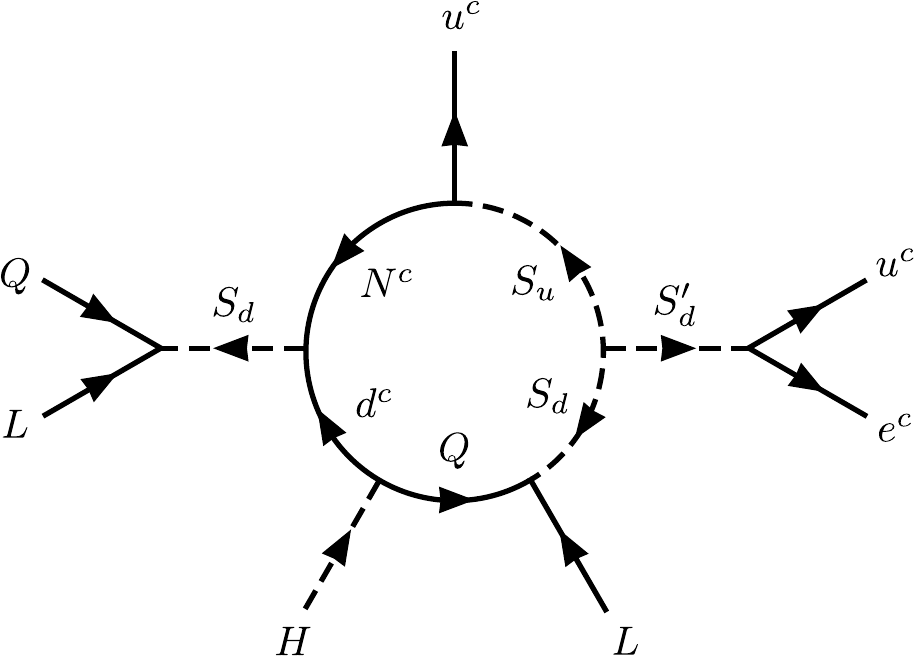}
\end{centering}
\protect\caption{\label{fig:loop_operator} Loop generation of the
  dimension 11 operator $\partial
  Q\overbar{u^{c}}\overbar{u^{c}}LL\overbar{e^{c}}H$ in our model. This leads to $p \to e^+ \overbar{\nu}\overbar{\nu} $, for example.}
	\end{figure}
	\par
\end{center}

With small changes to the model we have presented, it is also possible
to build the operator in eq. \eqref{eq:d13_1} involving only
right-handed fermions, instead of operator \eqref{eq:d13_2}. This
simply requires that $S_d$ and $S^\prime_d$ both have a coupling only to
$\overbar{e^c} \overbar{u^c}$ (and not $Q L$). The practical effect of this
change is that the proton decay mode into three charged leptons would
become the dominant mode.

\section{Comment on $\Delta L=3$, discrete symmetries, and Dirac neutrinos}
\label{sect:sym}

As we stated before, from simple dimensionality arguments, one expects
that, if allowed by symmetries, $\left(\Delta B,\Delta L\right)=\left(
1,\pm 1\right)$ processes will completely dominate over the 
higher dimensional $\left(\Delta B,\Delta L\right)=\left( 1,\pm
3\right)$ processes. Thus, proton decay in $\Delta L = 3$ modes will
be dominant only if the $\Delta L = 1$ mode is forbidden by some 
symmetry. As stated before, such a scenario can arise if the
accidental $U(1)_L$ (or $U(1)_{B+L}$) symmetry of Standard Model
is broken to a residual $Z_3$ (or $Z_4)$ subgroup by the new physics
involved in the process.

These symmetries force neutrinos to be Dirac particles, as Majorana
mass terms with $\left(\Delta B,\Delta L\right)=\left(0,2\right)$ are
forbidden. After electroweak symmetry breaking, the relevant mass
terms are $\left(Y_{\nu}\right)_{ik}\left\langle H\right\rangle
\nu_{i}N_{k}^{c}+\left(m_{N}\right)_{jk}N_{j}N_{k}^{c}+\textrm{h.c.}$,
where $i,j=1,2,3$ and $k=1,\cdots,6$. For
$Y_{\nu}m_{N}^{\dagger}\approx\boldsymbol{0}$, the 3 left-handed $\nu$
states will not mix significantly with $N^c$, and the light neutrino
masses and mixing angles will depend only on the matrix
$Y_{\nu}Y_{\nu}^{\dagger}\left\langle H\right\rangle ^{2}$.  It is
then possible to reproduce oscillation data with small $Y_{\nu}$
couplings. More elegant alternatives to this simple construction
require additional fields, which can be used to explain the smallness
of the observed neutrino masses through a Dirac seesaw mechanism
\cite{Roy:1983be,Chulia:2016ngi}.

Note that, in the absence of a $Z_3(L)$ symmetry (or an equivalent
one), it will be possible to use the Majorana neutrino mass operator
to convert $\left(\Delta B,\Delta L\right)=\left(1,\pm3\right)$
operators into the $\left(\Delta B,\Delta
L\right)=\left(1,\pm1\right)$ standard proton decay. However, due to
the smallness of neutrino masses, this does not imply that the 2-body
decay rate of the proton will necessarily be large.

Finally, we would like to mention that the $Z_3(L)$ is insufficient to
explain the absence of some couplings in our model. In particular, in
order to avoid meson decay constraints and still maintain a sizable
proton-decay rate, the two scalar $S_d$ and $S^\prime_d$ which share
the same quantum numbers must couple differently to $QL$ and $\overbar{
  e^c} \overbar{u^c}$. We assumed that $S_d$ couples mostly to the former
fermions only, and $S^\prime_d$ to the latter. This arrangement can be
achieved with the introduction of a second Higgs doublet field
$H^\prime$ and an extra $Z_2$ symmetry such that the charge of $e^c$,
$S^\prime_d$ and all $SU(2)_L$ doublets in the model is $-1$. In this
way, the terms $S^\prime_d Q L$ and $S_d \overbar{e^c} \overbar{u^c}$ are
forbidden. Furthermore, quarks and leptons will couple to $H$ and
$H^\prime$, respectively, hence both Higgs doublets need to acquire a
non-zero vacuum expectation value. This will spontaneous break the
$Z_2$ symmetry, generating $S^\prime_d Q L$ and $S_d \overbar{e^c} \overbar{
  u^c}$ couplings at loop level, which are not problematic
phenomenologically.

\section{Conclusions}
\label{sect:cncl}

Even if lepton number $L$ is not a conserved quantity, it is still
possible that all processes which violate it do so in multiples of
some number $n$. In this work we considered the possibility that
leptons can only be created or destroyed in units of $n=3$. This
implies that there is a remnant $Z_3(L)$ symmetry and hence neutrinos
are Dirac particles. The processes associated to this type of lepton
number violation do not conserve baryon number $B$ either. The case
$\Delta\left(B,L\right)=\left(1,3\right)$ which we considered is not only the simplest one, but it
also leads to nucleon decay into three leptons. Given the high
dimensionality of the relevant operators, proton and neutron decay
life-time bounds are satisfied even for TeV mediator masses.  There is
the interesting possibility that both Hyper-Kamiokande and the LHC
will be able to probe these scenarios.  We have suggested a particular
model implementing these ideas, where singlet fermions and three new
scalar lepto-quarks are added to the Standard Model. We showed by
choosing a benchmark scenario with 1 TeV scalar masses and benchmark
values of the couplings that such a scenario is feasible. In this case
the nucleon decay modes visible at Hyper-Kamiokande, and the LHC
signatures will depend on the mass of the singlet fermions. For a TeV
scale mass, it should be possible to observe events with 3 same sign
leptons plus jets and no missing energy at LHC. For lower masses,
proton decay into 4- and 5-body final state might be observable at DUNE or
Hyper-Kamiokande.

\begin{acknowledgments}

We would like to thank Ricardo Cepedello for helping us with the proton lifetime calculations, and also
Christoph Ternes for clarifying some details concerning the DUNE experiment. This work  was supported by the Spanish grants FPA2017-85216-P and SEV-2014-0398
(MINECO), and PROMETEOII/2014/084 (Generalitat Valenciana).  R.F. was
also financially supported through the grant Juan de la
Cierva-formaci\'on FJCI-2014-21651. R.S. would like to dedicate this
work to the memory of his friend Ravindra Singh Solanki.
	
\end{acknowledgments}



\begin{thebibliography}{10}
	
	\bibitem{Weinberg:1979sa}
	S.~Weinberg,
	\newblock Phys. Rev. Lett. {\bf 43}, 1566 (1979).
	
	\bibitem{Wilczek:1979hc}
	F.~Wilczek and A.~Zee,
	\newblock Phys. Rev. Lett. {\bf 43}, 1571 (1979).
	
	\bibitem{Abbott:1980zj}
	L.~F. Abbott and M.~B. Wise,
	\newblock Phys. Rev. {\bf D22}, 2208 (1980).
	
	\bibitem{Nath:2006ut}
	P.~Nath and P.~Fileviez~Perez,
	\newblock Phys. Rept. {\bf 441}, 191 (2007), arXiv:hep-ph/0601023.
	
	\bibitem{Patrignani:2016xqp}
	Particle Data Group, C.~Patrignani {\em et~al.},
	\newblock Chin. Phys. {\bf C40}, 100001 (2016).
	
	\bibitem{Avignone:2007fu}
	F.~T. Avignone, III, S.~R. Elliott, and J.~Engel,
	\newblock Rev. Mod. Phys. {\bf 80}, 481 (2008), arXiv:0708.1033.
	
	\bibitem{Deppisch:2012nb}
	F.~F. Deppisch, M.~Hirsch, and H.~Pas,
	\newblock J. Phys. {\bf G39}, 124007 (2012), arXiv:1208.0727.
	
	\bibitem{Hambye:2017qix}
	T.~Hambye and J.~Heeck,
	\newblock (2017), arXiv:1712.04871.
	
	\bibitem{Learned:1979gp}
	J.~Learned, F.~Reines, and A.~Soni,
	\newblock Phys. Rev. Lett. {\bf 43}, 907 (1979),
	\newblock [Erratum: Phys. Rev. Lett.43,1626(1979)].
	
	\bibitem{Reines:1974pb}
	F.~Reines and M.~F. Crouch,
	\newblock Phys. Rev. Lett. {\bf 32}, 493 (1974).
	
	\bibitem{Cherry:1981uq}
	M.~L. Cherry {\em et~al.},
	\newblock Phys. Rev. Lett. {\bf 47}, 1507 (1981).
	
	\bibitem{Takhistov:2014pfw}
	Super-Kamiokande, V.~Takhistov {\em et~al.},
	\newblock Phys. Rev. Lett. {\bf 113}, 101801 (2014), arXiv:1409.1947.
	
	\bibitem{Weinberg:1980bf}
	S.~Weinberg,
	\newblock Phys. Rev. {\bf D22}, 1694 (1980).
	
	\bibitem{Durieux:2012gj}
	G.~Durieux, J.-M. Gerard, F.~Maltoni, and C.~Smith,
	\newblock Phys. Lett. {\bf B721}, 82 (2013), arXiv:1210.6598.
	
	\bibitem{Kovalenko:2002eh}
	S.~Kovalenko and I.~Schmidt,
	\newblock Phys. Lett. {\bf B562}, 104 (2003), arXiv:hep-ph/0210187.
	
	\bibitem{Appelquist:2001mj}
	T.~Appelquist, B.~A. Dobrescu, E.~Ponton, and H.-U. Yee,
	\newblock Phys. Rev. Lett. {\bf 87}, 181802 (2001), arXiv:hep-ph/0107056.
	
	\bibitem{Babu:2003qh}
	K.~S. Babu, I.~Gogoladze, and K.~Wang,
	\newblock Phys. Lett. {\bf B570}, 32 (2003), arXiv:hep-ph/0306003.
	
	\bibitem{tHooft:1976rip}
	G.~'t~Hooft,
	\newblock Phys. Rev. Lett. {\bf 37}, 8 (1976).
	
	\bibitem{tHooft:1976snw}
	G.~'t~Hooft,
	\newblock Phys. Rev. {\bf D14}, 3432 (1976),
	\newblock [Erratum: Phys. Rev.D18,2199(1978)].
	
	\bibitem{Davidson:1993qk}
	S.~Davidson, D.~C. Bailey, and B.~A. Campbell,
	\newblock Z. Phys. {\bf C61}, 613 (1994), arXiv:hep-ph/9309310.
	
	\bibitem{Dorsner:2016wpm}
	I.~Doršner, S.~Fajfer, A.~Greljo, J.~F. Kamenik, and N.~Košnik,
	\newblock Phys. Rept. {\bf 641}, 1 (2016), arXiv:1603.04993.
	
	\bibitem{CMS:2016imw}
	CMS, C.~Collaboration,
	\newblock (2016).
	
	\bibitem{Aaboud:2016qeg}
	ATLAS, M.~Aaboud {\em et~al.},
	\newblock New J. Phys. {\bf 18}, 093016 (2016), arXiv:1605.06035.
	
	\bibitem{Aoki:2017puj}
	Y.~Aoki, T.~Izubuchi, E.~Shintani, and A.~Soni,
	\newblock Phys. Rev. {\bf D96}, 014506 (2017), arXiv:1705.01338.
	
	\bibitem{Claudson:1981gh}
	M.~Claudson, M.~B. Wise, and L.~J. Hall,
	\newblock Nucl. Phys. {\bf B195}, 297 (1982).
	
	\bibitem{Abe:2011ts}
	K.~Abe {\em et~al.},
	\newblock (2011), arXiv:1109.3262.
	
	\bibitem{Acciarri:2015uup}
	DUNE, R.~Acciarri {\em et~al.},
	\newblock (2015), arXiv:1512.06148.
	
	\bibitem{Dorsner:2018ynv}
	I.~Doršner and A.~Greljo,
	\newblock (2018), arXiv:1801.07641.
	
	\bibitem{Roy:1983be}
	P.~Roy and O.~U. Shanker,
	\newblock Phys. Rev. Lett. {\bf 52}, 713 (1984),
	\newblock [Erratum: Phys. Rev. Lett.52,2190(1984)].
	
	\bibitem{Chulia:2016ngi}
	S.~Centelles~Chuliá, E.~Ma, R.~Srivastava, and J.~W.~F. Valle,
	\newblock Phys. Lett. {\bf B767}, 209 (2017), arXiv:1606.04543.
	
\end{thebibliography}
%
%

\end{document}